\documentclass[]{interact}

\usepackage{epstopdf}
\usepackage{xcolor,soul}
\usepackage[caption=false]{subfig}

\usepackage[numbers,sort&compress]{natbib}
\bibpunct[, ]{[}{]}{,}{n}{,}{,}
\renewcommand\bibfont{\fontsize{10}{12}\selectfont}
\makeatletter
\def\NAT@def@citea{\def\@citea{\NAT@separator}}
\makeatother

\theoremstyle{plain}

\theoremstyle{definition}

\theoremstyle{remark}

\begin{document}

\articletype{RESEARCH ARTICLE}

\title{Multimode Interferometers: an Analytical Method for Determining the Accumulated Phase Difference Between the Fundamental Mode and One Arbitrary High-Order Mode}

\author{
\name{Yuri Hayashi Isayama\textsuperscript{a}\thanks{CONTACT Y. H. Isayama. Email: yuri-isayama@ufmg.br}}
\affil{\textsuperscript{a} Department of Physics, Federal University of Minas Gerais, Av. Presidente Antônio Carlos, 6627, Belo Horizonte, Minas Gerais, Brazil.}
}

\maketitle

\begin{abstract}
In multimode interferometers, the interaction between several modes brings a high level of complexity to the interpretation of its light patterns. With recent advances, it is possible to selectively excite only a couple of modes inside the device. This paper presents an analytical method for determining the phase difference between two propagating modes in the multimode interferometer, in an effort to simplify the usage and expand the range of applications for this type of optical devices.
\end{abstract}


\section{Introduction}
\label{sec:intro}
Interferometers represent one fundamental building block in the field of integrated optics. Its applications can vary in a wide range, such as switches and modulators \cite{2022_babaki}, optical logic gates \cite{2021_xiao}, and optical sensors \cite{2022_noman,2022_laplatine,2019_jhonattan}. One type of interferometer is the Multimode Interferometer (MMI). In this structure, light is distributed among different optical modes that travel with different propagation constants. At the end of the MMI, the light pattern observed depends on the different phases accumulated by the modes throughout the propagation distance. This working mechanism has been widely used in recent years in sensing applications \cite{2016_hoppe,2017_gonzalez,2018_dwivedi,2019_grajales}. The sensitivity of the device has been improved through different approaches, such as: coupling structures to improve light coupling to the optical modes \cite{2019_grajales,2019_ebihara,2018_liang}, engineering of the waveguide core to enhance the interaction between light and matter \cite{2021_moran},and the use of high-order modes \cite{2015_jhonattan,2021_isayama,2024_isayama} and different polarizations (TE/TM) \cite{2023_isayama,2023_isayama_SI}. The introduction of higher-order modes can drastically improve device sensitivity, but the interaction between the modes also becomes much more complex. One solution is to reduce the number of propagating modes within the MMI with new coupling structures, as presented in \cite{2021_isayama}, but the interpretation of the output light is not straightforward. In this paper, an analytical method for determining the phase difference between the fundamental mode (both TE or TM) and an arbitrary high-order (order $n$) mode at the end of the MMI is presented.

\section{Mathematical Modeling}

A MMI is basically composed of three sections: an input section, which is responsible for exciting the different propagating modes within the MMI; a multimode waveguide (MMW), where all the excited modes propagate through a certain distance; and an output section, where light is captured by a physical structure (such as another waveguide) or a detection instrument (photodetector, photodetector array, CCD camera, etc.). The analysis here presented concerns the multimode waveguide (MMW) section of the MMI.

\subsection{Basic expression for optical power at the end of the MMW}
\begin{figure}[ht]
\centering
\fbox{\includegraphics[width=\linewidth]{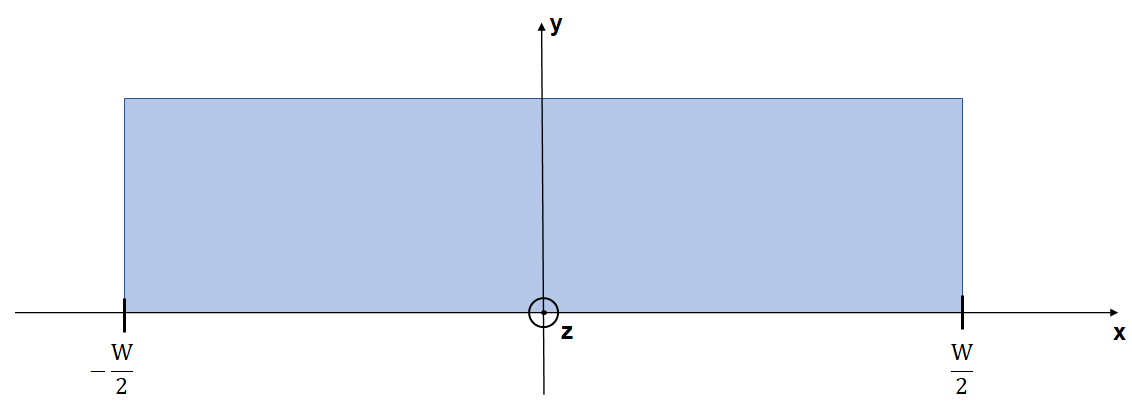}}
\caption{Cross-section of a multimode waveguide.}
\label{fig:MMW-cross}
\end{figure}

Consider the multimode waveguide (MMW) section of a Multimode Interferometer (MMI) with a cross-section as shown in Figure \ref{fig:MMW-cross}. Inside the MMW (i.e. $|x| \leq W/2$), the $m^{\mathrm{th}}$ Transverse Electric (TE) mode has the following electric field distribution:

\begin{equation}
    E_x^m (x,z) = A_m cos\left( \frac{2u_m}{W}x - \frac{m\pi}{2}\right)e^{-j\beta_mz},
    \label{eq:1}
\end{equation}
where $E_x$ denotes the electric field in the $x$ direction, $m$ is the number of the mode ($m=0$ represents the fundamental TE$_0$ mode, $m=1$ the TE$_1$ mode, and so on), $A_m$ is the amplitude of the $m^{\mathrm{th}}$ mode, $W$ is the MMW core width, $\beta_m$ is the propagation constant of the $m^{\mathrm{th}}$ mode, and u$_m \approx (m+1)\frac{\pi}{2}$ \cite{okamoto}.

If there are solely two propagating modes in the MMW, TE$_0$ and TE$_n$, at the end of the MMW (i.e. $z = L_{MMI}$), the electric field distribution will be as follows:

\begin{multline}
    E_x(x,L_{MMI}) = A_0 cos\left( \frac{\pi}{W}x\right)e^{-j\beta_0L_{MMI}} + A_n cos\left( \frac{(n+1)\pi}{W}x-\frac{n\pi}{2}\right)e^{-j\beta_nL_{MMI}}.
\label{eq:2}
\end{multline}

The optical power is proportional to $|E_x^2|$, therefore

\begin{multline}
    \lvert E_x(x,L_{MMI})\rvert^2 = \bigg| A_0 cos\left(\frac{\pi}{W}x\right)e^{-j\phi_0} + A_n cos\left(\frac{(n+1)\pi}{W}x - \frac{n\pi}{2} \right)e^{-j\phi_n} \bigg|^2,
    \label{eq:3}
\end{multline}
with $\phi_\alpha = \beta_\alpha L_{MMI}$. Applying the law of cosines to expand Eq. \ref{eq:3}, we obtain


\begin{multline}
    \left|E_x(x,L_{MMI})\right|^2 = A_0^2cos^2\left(\frac{\pi}{W}x\right) + A_n^2cos^2\left(\frac{(n+1)\pi}{W}x-\frac{n\pi}{2}\right) - \\
    - 2A_0A_n cos\left(\frac{\pi}{W}x\right) cos\left(\frac{(n+1)}\pi{W}x - \frac{n\pi}{2} \right)cos(\phi_0 - \phi_n).
    \label{eq:4}
\end{multline}

The TE$_n$ mode presents maximum electric field intensity at the points

\begin{equation*}
    \frac{(n+1)\pi}{W}x - \frac{n\pi}{2} = \tau\pi, \tau \in \mathbb{Z}^{-}\cup\{0\}
\end{equation*}
\begin{equation}
    x = \left(\tau + \frac{n}{2} \right)\frac{W}{n+1}, |x| \leq \frac{W}{2}.
    \label{eq:5}
\end{equation}

In each of these points, the optical power is given by:

\begin{multline}
    \left|E_x\left(x = (\tau+\frac{n}{2})\frac{W}{n+1}, z = L_{MMI}\right) \right|^2 = \\
    = A_0^2 cos^2\left[(\tau+\frac{n}{2})\frac{\pi}{n+1} \right] + A_n^2 cos^2\left[(\tau + \frac{n}{2})\pi - \frac{n}{2}\pi \right] - \\
    - 2A_0A_n cos\left[(\tau+\frac{n}{2})\frac{\pi}{n+1}\right] cos\left[(\tau+\frac{\pi}{2})\pi - \frac{n\pi}{2} \right] cos\Delta\phi,
    \label{eq:6}
\end{multline}
with $\Delta\phi = \phi_0 - \phi_n$. Simplifying the arguments of cosines (and because $cos^2[\tau\pi] = 1$), we arrive at 

\begin{multline}
    \left|E_x\left((\tau+\frac{n}{2})\frac{W}{n+1},L_{MMI}) \right) \right|^2 = A_0^2 cos^2\left(\frac{\tau+\frac{n}{2}}{n+1}\pi\right) + \\
    + A_n^2 - 2A_0A_n cos\left(\frac{\tau+\frac{n}{2}}{n+1}\pi\right)cos(\tau\pi) cos\Delta\phi.
    \label{eq:8}
\end{multline}

\subsection{Some remarks on number of optical power peaks and their location}

Assuming the following hypothesis: the number of optical power maxima is always equal to n+1. We defined $\tau \in \mathbb{Z}^-\cup\{0\}$ and $|x|<\frac{W}{2}$. If there were more optical maxima than n+1, then there should be a value of $\tau$ such as $\tau \leq -(n+1)$. The optical power maxima occur in the points

\begin{equation}
    |x| = \left|\left(\tau+\frac{n}{2}\right)\frac{W}{n+1} \right| = \left|\tau+\frac{n}{2}\frac{W}{n+1}\right|,
    \label{eq:p1}
\end{equation}
since $n \geq 0$. If $\tau = -(n+1)$, then

\begin{equation}
    \left|\tau + \frac{n}{2} \right| = -\left( \tau +\frac{n}{2}\right).
    \label{eq:p2}
\end{equation}

The power peak location corresponding to $\tau = -(n+1)$ is

\begin{equation}
    |x| = - \left(\frac{-2(n+1)}{n+1} \right)\frac{W}{2} = W > \frac{W}{2}.
    \label{eq:p3}
\end{equation}
Therefore, $\tau < -(n+1)$, $|x|> \frac{W}{2}$ and, thus, the number of maxima will always be n+1, with $\tau \in \{-n, -n+1, ..., 1, 0\}$.

Because of the symmetry of the MMW core, it is expected that the optical power peaks are located in symmetrical positions in relation to the plane $x = 0$. If there is a peak in $x = x_0$, there should be a symmetrical peak in $x = -x_0$ (except for the case of $x_0 = 0$, when there is a peak at the origin, which means the optical mode has even symmetry and an odd number of peaks -- i.e. $n = 2k+1$, $k \in \mathbb{Z}$).

Consider a MMW with a mode of order $n$ (TE$_n$) and the pair of cases $\tau = 0$ and $\tau = -n$:

\begin{equation*}
        \tau = 0 \Rightarrow x = \left(\frac{n}{2} \right)\frac{W}{n+1}=\frac{n}{n+1}\frac{W}{2}=x_0
\end{equation*}
\begin{equation*}
        \tau = -n \Rightarrow x = \left(-n + \frac{n}{2} \right)\frac{W}{n+1}=\frac{-n}{n+1}\frac{W}{2}=x_n
\end{equation*}
which gives us $x_0 = x_n$. 

If we expand this to the case $\tau = -k$ and $\tau = -(n-k)$:

\begin{equation}
        \tau = -k \Rightarrow x = \left(-k + \frac{n}{2} \right)\frac{W}{n+1}=\left(\frac{-2k}{n+1}+\frac{n}{n+1}\right)\frac{W}{2}=x_k
        \label{eq:p4}
\end{equation}
\begin{multline}
        \tau = -(n-k) \Rightarrow \\
        x = \left(-n+k + \frac{n}{2} \right)\frac{W}{n+1}=\left(\frac{2k}{n+1}-\frac{n}{n+1}\right)\frac{W}{2}=x_{n-k}.
        \label{eq:p5}
\end{multline}
The optical power peaks occur in pairs ($x_k$ and $x_{n-k}$) in terms of their position in the $x$ axis, where for TE$_{n}$ modes with even order ($n = 2\alpha$), there is an additional central peak. In the modes TE$_{n}$ with odd order ($n = 2\alpha +1$), this central peak does not exist.

\subsection{Developing an expression for the phase difference between the two propagating modes}

Let us revisit Eq. \ref{eq:3}:

\begin{multline*}
    \lvert E_x(x,L_{MMI})\rvert^2 = \bigg| A_0 cos\left(\frac{\pi}{W}x\right)e^{-j\phi_0} + A_n cos\left(\frac{(n+1)\pi}{W}x - \frac{n\pi}{2} \right)e^{-j\phi_n} \bigg|^2.
\end{multline*}
At the optical power peaks corresponding to $x = (\tau + \frac{n}{2})\frac{W}{n+1}$, the argument of the second cosine in Eq. \ref{eq:3} shall assume values of $\frac{(n+1)\pi}{W}x - n\frac{\pi}{2} = \tau\pi$, with $\tau \in \mathbb{Z}^- \cup\{0\}$. Starting from $\tau = 0$ to $\tau = -n$, one may note that $cos\left[\frac{(n+1)pi}{W}x-n\frac{\pi}{2}\right] = cos(\tau\pi)$. The value of the cosine alternates between $cos(0)$ and $cos(\pi)$ for the different values of $\tau$. The result is that for consecutive peaks, the value of $cos\left[\frac{(n+1)pi}{W}x-n\frac{\pi}{2}\right]$ alternates between 0 and -1, and if $x_\tau = \left(\tau + \frac{n}{2} \right)\frac{W}{n+1}$, then we may say that

\begin{multline}
    |E_x(x_\tau,L_{MMI})|^2 = \bigg|A_0cos\left(\frac{\pi}{W}x_\tau\right)e^{-j\phi_0} + A_n cos\left(\frac{n+1}{W}x_\tau - n\frac{\pi}{2}\right)e^{-j\phi_n}\bigg|^2 = \\
    = A_0^2 cos^2\left(\frac{\pi}{W}x_\tau\right) + A_n^2 - 2A_0A_n cos\left(\frac{\pi}{W}x_\tau \right) cos(\tau\pi) cos\Delta\phi
    \label{eq:9}
\end{multline}
and for $x = x_{\tau-1}$

\begin{multline}
   |E_x(x_{\tau-1},L_{MMI})|^2 = A_0^2 cos^2\left(\frac{\pi}{W}x_{\tau-1}\right) + A_n^2 - 2A_0A_n cos\left(\frac{\pi}{W}x_{\tau-1}\right) cos(\tau\pi) cos(\Delta\phi).
    \label{eq:10}
\end{multline}
There are two cases that need to be analyzed: when the order $n$ is odd and when it is even.

If $n$ is odd: there is an even number of power peaks equal to n+1. If $k$ is a natural number, then every peak represented by $\tau = k$ has its symmetrical $\tau = -(n-k)$. These symmetrical peaks are such that the term $2A_0A_n cos\left(\frac{\pi}{W}x_k\right)cos(\tau\pi)cos\Delta\phi$ equals to $2A_0A_n cos\left(\frac{\pi}{W}x_{n-k}\right)cos(\tau\pi)cos\Delta\phi$, since $x_k = -x_{n-k}$ (see Eqs. \ref{eq:p4} and \ref{eq:p5}). Because the signal of such term is inverted for consecutive values of $\tau$ (see Eqs. \ref{eq:9} and \ref{eq:10}), the existence of an even number of power peaks implies that the terms $2A_0A_n cos\left(\frac{\pi}{W}x_k\right)cos(\tau\pi)cos\Delta\phi$ and $2A_0A_n cos\left(\frac{\pi}{W}x_{n-k}\right)cos(\tau\pi)cos\Delta\phi$ will present opposite signs and, thus, we have

\begin{equation}
    |E_x(x_k,L_{MMI})|^2 + |E_x(x_{n-k},L_{MMI})|^2 = 2A_0^2 cos^2\left(\frac{\pi}{W}x_k \right) + 2A_n^2
    \label{eq:11}
\end{equation}
If we add all the power peaks, the result will be

\begin{equation}
    \sum\limits_{\tau=0}^{-n}  |E_x(x_\tau,L_{MMI})|^2 = \sum\limits_{\tau=0}^{-n} \left[A_0^2 cos^2\left(\frac{\pi}{W}x_\tau \right) + A_n^2\right], 
    \label{eq:12}
\end{equation}
which is independent from the value of $\Delta\phi$. On the other hand, if we subtract the power of symmetrical peaks:

\begin{multline}
    |E_x(x_k,L_{MMI})|^2 - |E_x(x_{n-k},L_{MMI})|^2 = \\
    = -4A_0A_n cos\left(\frac{\pi}{W}x_k \right) cos(k\pi)cos\Delta\phi.
    \label{eq:13}
\end{multline}
Performing the same sum of all pairs of power peaks:

\begin{multline}
    \sum\limits_{k=1}^{\frac{n-1}{2}}  |E_x(x_k,L_{MMI})|^2 - |E_x(x_{n-k},L_{MMI})|^2 = \\
    = \sum\limits_{k=1}^{\frac{n-1}{2}} -4A_0A_n cos\left(\frac{\pi}{W}x_k\right)cos(k\pi)cos\Delta\phi.
    \label{eq:14}
\end{multline}
Finally, let us define the adimensional parameter $I_{odd}$ as:

\begin{equation*}
    I_{odd} = \frac{\sum\limits_{k=1}^{\frac{n-1}{2}}  |E_x(x_k,L_{MMI})|^2 - |E_x(x_{n-k},L_{MMI})|^2}{\sum\limits_{\tau=0}^{-n}  |E_x(x_\tau,L_{MMI})|^2} \Rightarrow
\end{equation*}
\begin{equation}
   \Rightarrow I_{odd} = \frac{\sum\limits_{k=1}^{\frac{n-1}{2}} -4A_0A_n cos\left(\frac{\pi}{W}x_k\right)cos(k\pi)}{\sum\limits_{\tau=0}^{-n} \left[A_0^2 cos^2\left(\frac{\pi}{W}x_\tau \right) + A_n^2\right]}cos\Delta\phi.
    \label{eq:15}
\end{equation}    

\begin{figure}[ht]
\centering
\fbox{\includegraphics[width=\linewidth]{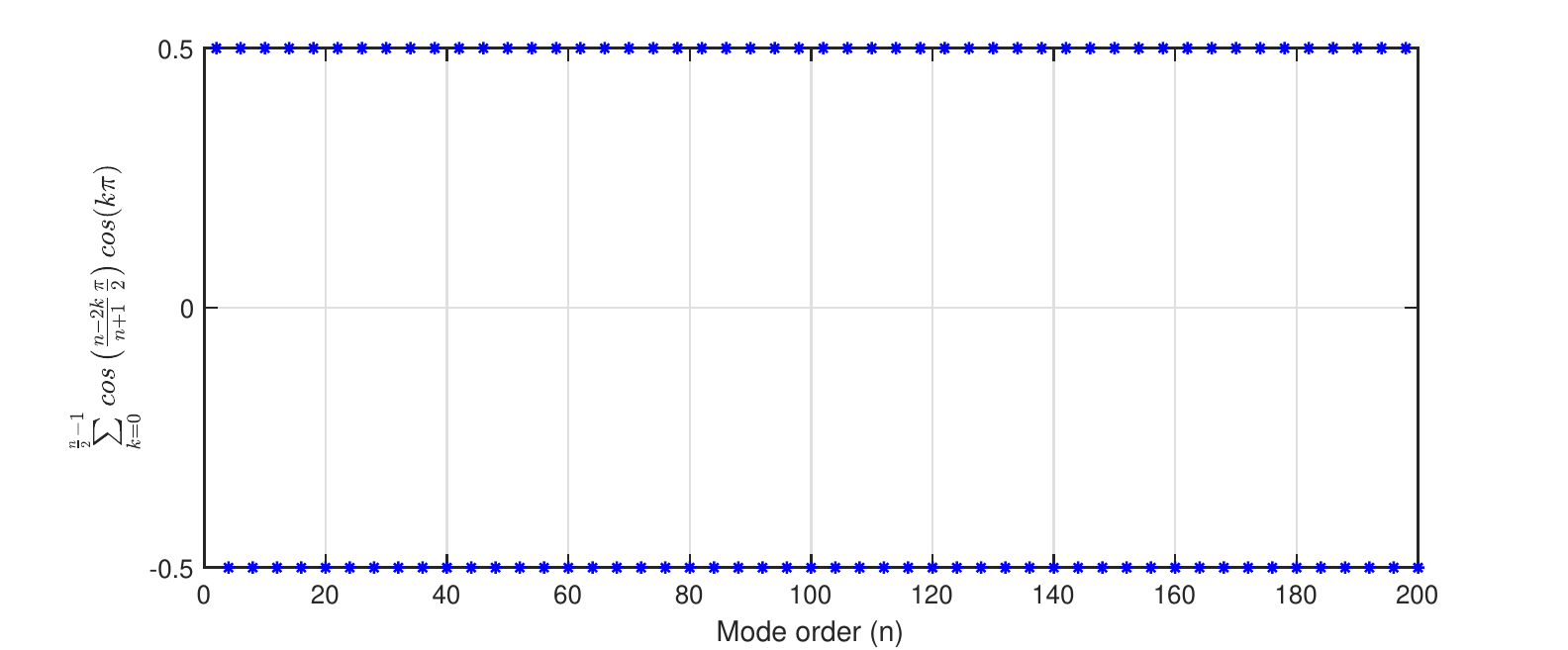}}
\caption{Calculation of the sum in Eq. \ref{eq:18}. The values of $n$ are all even non-negative integers. The result alternates between two values: $\frac{1}{2}$ and $-\frac{1}{2}$.}
\label{fig:identity}
\end{figure}

If $n$ is, now, an even number: there is an odd number of power peaks equal to n+1. The power peak corresponding to $\tau = -\frac{n}{2}$ is the central peak at the origin ($x = 0$) and it is the only one that does not possess a symmetrical one. As before, we may write

\begin{equation*}
    2A_0A_n cos \left(\frac{\pi}{W}x_k\right)cos\Delta\phi = 2A_0A_n cos\left(\frac{\pi}{W}x_{n-k}\right)cos\Delta\phi
\end{equation*}
but in the case of $n$ even, these terms will present the same sign:

\begin{multline}
    |E_x(x_k,L_{MMI})|^2 + |E_x(x_{n-k},L_{MMI})|^2 = \\
    = 2A_0^2 cos^2\left(\frac{\pi}{W}x_k\right) + 2A_n^2 - 4A_0A_n cos\left(\frac{\pi}{W}x_{n-k}\right) cos(k\pi) cos\Delta\phi
    \label{eq:16}
\end{multline}
Adding all the peaks, including the central one, we have

\begin{multline}
    \sum\limits_{\tau=0}^{-n} |E_x(x_\tau,L_{MMI})|^2 = \\
    = A_0^2+A_n^2 - 2A_0A_n cos\left(\frac{n}{2}\pi\right)cos\Delta\phi + \\
    + \sum\limits_{k=0}^{\frac{n}{2}-1}\left[2A_0 cos^2\left(\frac{\pi}{W}x_k\right) + 2A_n^2 - 4A_0A_n cos\left(\frac{\pi}{W}x_k \right) cos(k\pi) cos\Delta\phi \right].
    \label{eq:17}
\end{multline}
The arguments $\frac{\pi}{W}x_k$ will have the form:

\begin{equation*}
    \frac{\pi}{W}x_k = \frac{\pi}{W}\left(-k+\frac{n}{2}\right)\frac{W}{n+1} = \frac{n-2k}{n+1}\frac{\pi}{2}.
\end{equation*}
The sum of cosines in Eq. \ref{eq:17}, $\sum\limits_{k=0}^{\frac{n}{2}-1} cos\left(\frac{\pi}{W}x_k\right) cos(k\pi)$ has an interesting property. Figure \ref{fig:identity} shows that the result of this sum alternates between two values, depending on $n$: $\frac{1}{2}$ and $-\frac{1}{2}$, and we may write

\begin{multline}
    \sum\limits_{k=0}^{\frac{n}{2}-1} cos\left(\frac{\pi}{W}x_k\right) cos(k\pi) = \sum\limits_{k=0}^{\frac{n}{2}-1} cos\left(\frac{n-2k}{n+1}\frac{\pi}{2}\right) cos(k\pi) = \\
    = \frac{(-1)^{\frac{n}{2}}}{2} = \frac{1}{2}cos\left(\frac{n}{2}\pi\right)
    \label{eq:18}
\end{multline}

Substituting the result in Eq. \ref{eq:18} in Eq. \ref{eq:17}, the result is

\begin{multline}
    \sum\limits_{\tau=0}^{-n} |E_x(x_\tau,L_{MMI})|^2 = A_0^2+A_n^2 +  \sum\limits_{k=0}^{\frac{n}{2}-1}\left[2A_0^2 cos^2\left(\frac{\pi}{W}x_k\right)+2A_n^2 \right],
    \label{eq:19}
\end{multline}
which is independent of $cos\Delta\phi$. Ignoring the central power peak, represented by $x_\psi$, and adding the remaining peaks in the following manner: 

\begin{multline*}
    |E_x(x_0,L{MMI})|^2 - |E_x(x_1,L{MMI})|^2 + |E_x(x_2,L{MMI})|^2 - \\
    - ... + |E_x(x_n,L{MMI})|^2 = \sum\limits_{\alpha=0}^{-n} (-1)^\alpha|E_x(x_{\alpha},L_{MMI})|^2,
\end{multline*}
with $\alpha \in \mathbb{Z}^{-}+\{0\}-\{x_\psi\}$, results in

\begin{multline}
    |E_x(x_{\alpha},L_{MMI})|^2 = \sum\limits_{\alpha=0}^{-n} (-1)^\alpha \left[A_0^2 cos^2\left(\frac{\pi}{W}x_{\alpha}\right)+A_n^2 \right] - \\
    - 2A_0A_n cos\left(\frac{\pi}{W}x_{\alpha}\right) cos\Delta\phi,
    \label{eq:20}
\end{multline}
which is proportional to $cos\Delta\phi$. Let us, then, define the adimensional parameter $I_{even}$ as:

\begin{equation*}
    I_{even} = \frac{\sum\limits_{\alpha=0}^{-n} (-1)^\alpha|E_x(x_{\alpha},L_{MMI})|^2}{\sum\limits_{\tau=0}^{-n} |E_x(x_\tau,L_{MMI})|^2} \Rightarrow
\end{equation*}
\begin{equation}
   I_{even} = \frac{\sum\limits_{\alpha=0}^{-n} (-1)^\alpha \left[A_0^2 cos^2\left(\frac{\pi}{W}x_{\alpha}\right)+A_n^2 \right] - 2A_0A_n cos\left(\frac{\pi}{W}x_{\alpha}\right) cos\Delta\phi}{A_0^2+A_n^2 +  \sum\limits_{k=0}^{\frac{n}{2}-1}\left[2A_0^2 cos^2\left(\frac{\pi}{W}x_k\right)+2A_n^2 \right]}
    \label{eq:21}
\end{equation}  

Eqs. \ref{eq:15} and \ref{eq:21} provide expressions that are proportional to $cos\Delta\phi$ and allow us to infer the accumulated phase difference between the two modes within the MMI after an arbitrary propagating distance $L_{MMI}$. One important property of the adimensional parameters $I_{odd}$ and $I_{even}$, given respectively by Eqs. \ref{eq:15} and \ref{eq:21}, is that they are independent of the input power coupled to the MMI device. Denoting the input power coupled to the MMI as $A$, both $A_0$ and $A_n$ are related to it, depending on the excitation method employed in the input section of the MMI, and they respect the following relations:

\begin{equation}
    \begin{cases}
        A_0 = p_0A \\
        A_n = p_nA \\
        p_0 + p_n = 1
    \end{cases}
    \label{eq:A}
\end{equation}
Both $p_0$ and $p_n$ are constants and depend solely on the excitation method, assuming it is invariant with time. Substituting Eq. \ref{eq:A} in Eqs. \ref{eq:15} and \ref{eq:21}, we have

\begin{equation}
    I_{odd} = \frac{\sum\limits_{k=1}^{\frac{n-1}{2}} -4p_0p_n cos\left(\frac{\pi}{W}x_k\right)cos(k\pi)}{\sum\limits_{\tau=0}^{-n} \left[p_0^2 cos^2\left(\frac{\pi}{W}x_\tau \right) + p_n^2\right]}cos\Delta\phi,
    \label{eq:22}
\end{equation}
\begin{equation}
    I_{even} = \frac{\sum\limits_{\alpha=0}^{-n} (-1)^\alpha \left[p_0^2 cos^2\left(\frac{\pi}{W}x_{\alpha}\right)+p_n^2 \right] - 2p_0p_n cos\left(\frac{\pi}{W}x_{\alpha}\right) cos\Delta\phi}{p_0^2+p_n^2 +  \sum\limits_{k=0}^{\frac{n}{2}-1}\left[2p_0^2 cos^2\left(\frac{\pi}{W}x_k\right)+2p_n^2 \right]}.
    \label{eq:23}
\end{equation}
Both parameters, $I_{odd}$ and $I_{even}$ are independent of the input power, $A$, and suffer no influence from fluctuations in it. It is also important to note that the same analysis is valid for TM modes, as one only needs to exchange the $E_x$ for the $E_y$ field in Eq. \ref{eq:1} and the demonstration is straightforward.

\section{Examples of application of the method}

In sensing applications, for instance, the phase difference between the two propagating modes is an information often used to infer the characteristics of a solution or analyte of interest \cite{2017_gonzalez,2019_grajales,2019_jhonattan,2021_isayama}. To illustrate how the method can be employed, two examples are presented: one with a Bimodal Waveguide (BiMW) and one with a Trimodal Waveguide (TriMW). In both cases, we assume the excitation of only two modes, the fundamental and the highest order one.

\subsection{Bimodal Waveguide - Odd symmetry case}

\begin{figure}[ht]
\centering
\fbox{\includegraphics[width=\linewidth]{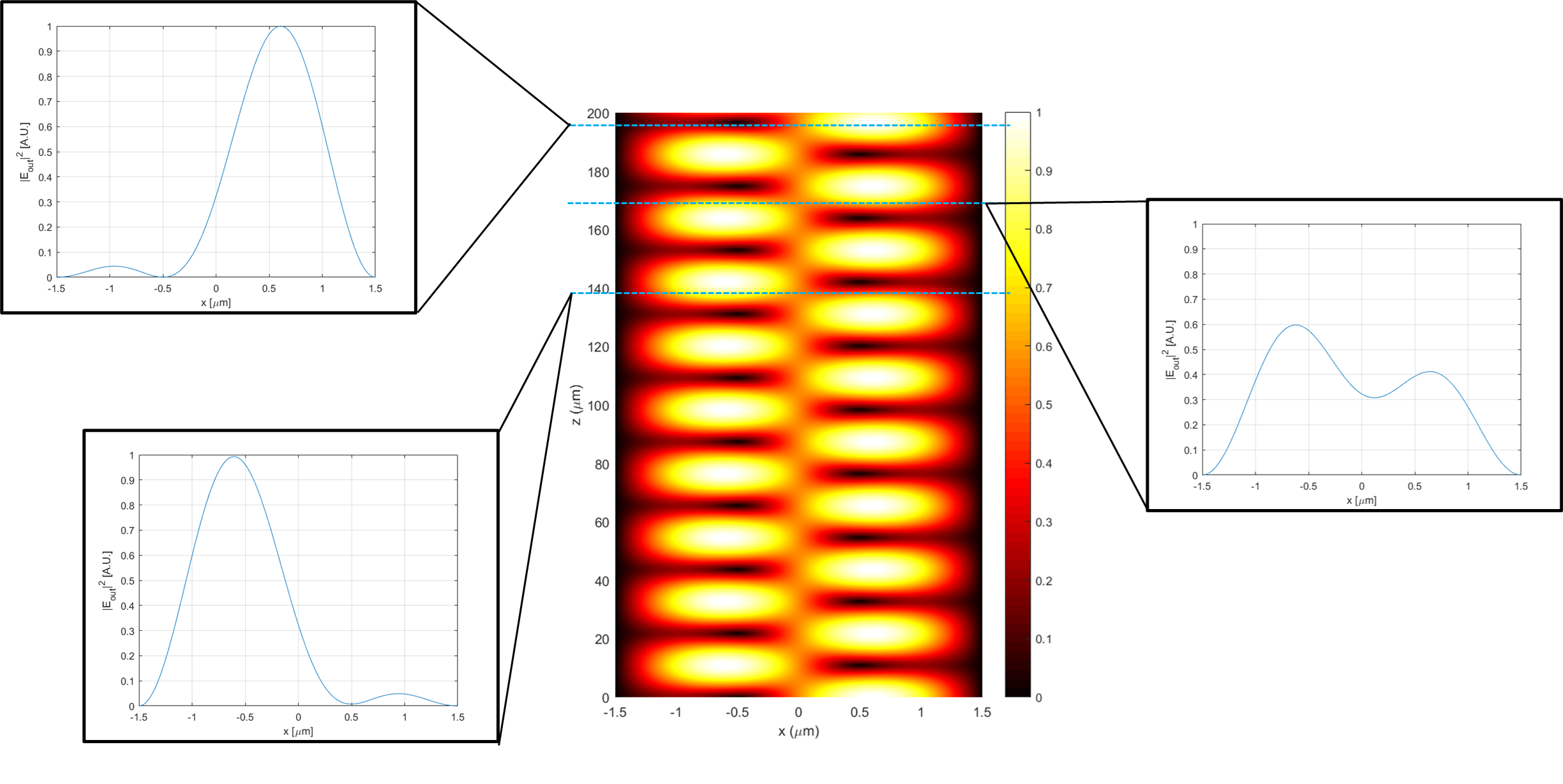}}
\caption{Light propagation within a BiMW with $L_{MMI} = 200 ~ \mu m$, $W = 3 ~\mu m$, $n_1 = 1.47$, and operating wavelength $\lambda = 1550$ nm. The excerpts show the field amplitude at different distances of the waveguide.}
\label{fig:BiWG-fields}
\end{figure}

In the Bimodal Waveguide (BiMW), the two propagating modes are the fundamental and the first order ones. The total electric field inside the BiMW can be calculated by Eq. \ref{eq:2} given a few parameters of the waveguide: the core width ($W$), the propagation constants of the two modes ($\beta_0$ and $\beta_1$), and the amplitude of each mode ($A_0$ and $A_1$). According to Eqs. \ref{eq:22} and \ref{eq:23}, the actual amplitude of each mode in not necessary, but only the proportion of the input field coupled to each ($p_0$ and $p_1$). If we substitute $L_{MMI}$ in Eq. \ref{eq:2} by $z$, then the expression give the total electric field as a function of the distance $z$. Figure \ref{fig:BiWG-fields} shows the propagation of the electric field inside the BiMW and the excerpts present the electric field amplitude for a few regions of interest. The BiMW has $W = 3 \mu m$, operating wavelength ($\lambda$) of 1550 nm, core refractive index of 1.47, and it is assumed that the amplitude of both modes is the same. The propagating constants can be calculated by \cite{okamoto}:

\begin{equation}
    \beta_m = \sqrt{k^2n_1^2 - (2u_m/W)^2},
    \label{eq:beta}
\end{equation}
where $k = 2\pi/\lambda$ and $n_1$ is the refractive index of the waveguide core.

\begin{figure}[ht]
\centering
\fbox{\includegraphics[width=4in]{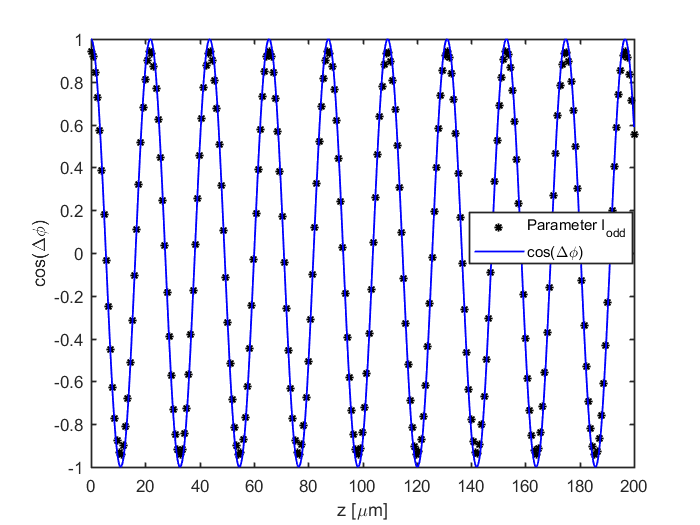}}
\caption{Comparison of the cosine of the actual phase difference ($\Delta\phi$) between the modes in the BiMW and the calculated parameter $I_{odd}$.}
\label{fig:BiWG-I}
\end{figure}

In a practical application, to determine the phase difference between the modes, one would have to measure the intensity of the electric field in the positions given by Eq. \ref{eq:5}. Since the highest order mode is the TE$_1$, we fall in the case where $n$ is an odd number and we utilize Eq. \ref{eq:15} to calculate the adimensional parameter $I_{odd}$. It is important to remember that $I_{odd}$ provides a value that is proportional to the cosine of the phase difference between the mode at the distance where the e-field intensities have been measured. Figure \ref{fig:BiWG-I} compares the value of $I_{odd}$ as a function of the distance with the actual phase difference between the modes (represented by $cos(\Delta\phi)$). It is evident that, although they might not be exactly the same, the curves are proportional and the information of $\Delta\phi$ can be obtained from the adimensional parameter proposed by Eq. \ref{eq:15}.

In the context of optical sensing, the idea is not to measure the electric field intensity at several points inside the MMI, but to measure at the end of the waveguide (at the distance $z = L_{MMI}$). Before the test subject (which can be a solution, analyte, temperature, etc.) is inserted in the sensing area, the phase difference between the modes will have a certain value. After the introduction of the test subject, the power measurements will change due to the perturbation of the propagating modes, rendering a new phase difference between them, which, in turn, is relatable to the change in the refractive index in the sensing area.

\subsection{Trimodal Waveguide - Even symmetry case}

\begin{figure}[ht]
\centering
\fbox{\includegraphics[width=\linewidth]{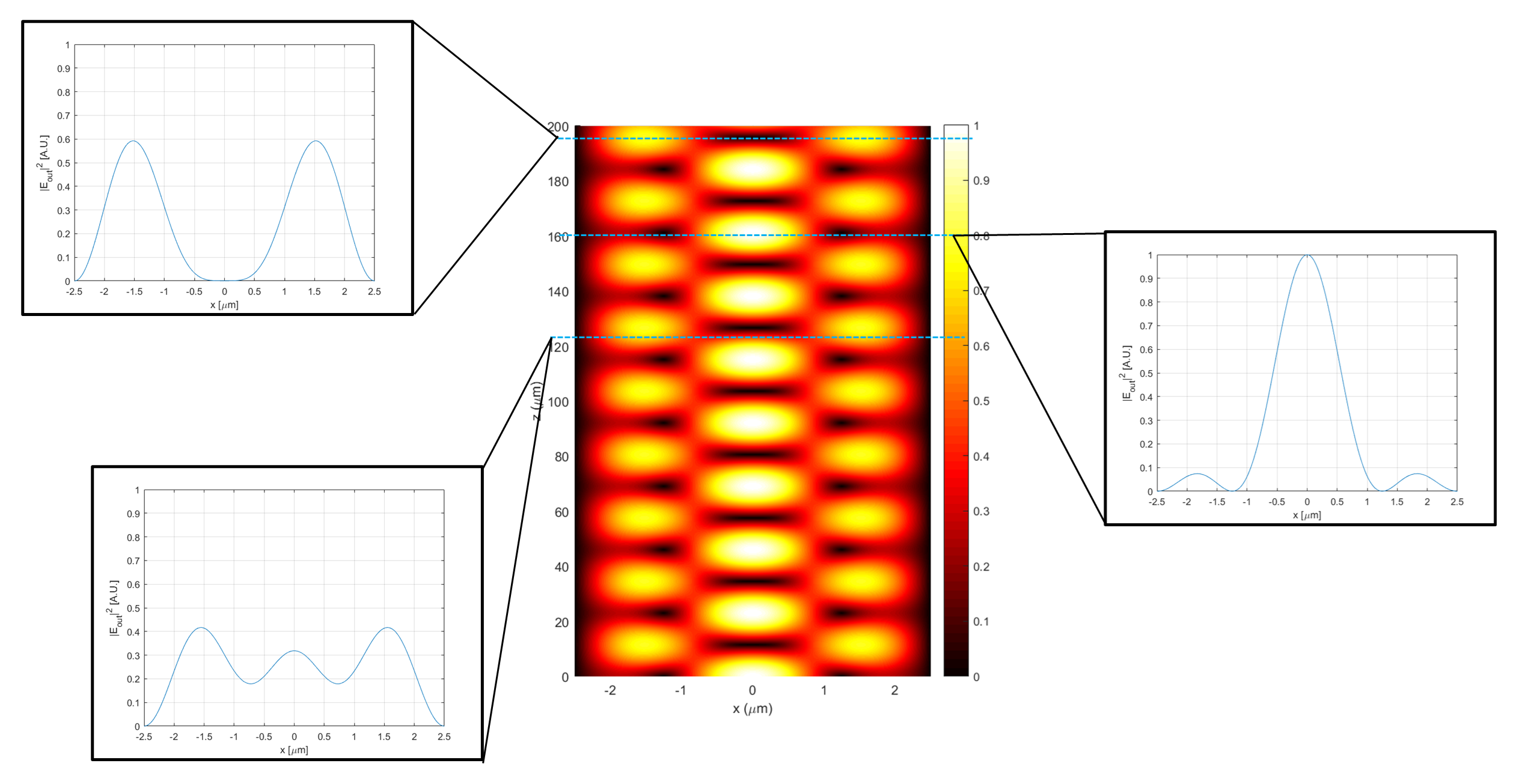}}
\caption{Light propagation within a TriMW with $L_{MMI} = 200 \mu m$, $W = 5 \mu m$, $n_1 = 1.47$, and operating wavelength $\lambda = 1550$ nm. The excerpts show the field amplitude at different distances of the waveguide.}
\label{fig:TriWG-fields}
\end{figure}

To exemplify the application of the method in a case with even symmetry, a Trimodal Waveguide (TriMW) is presented. Using the same methodology employed for the BiMW, the light propagation in a TriMW with $W = 5 ~ \mu m$, $n_1 = 1.47$ and $\lambda = 1550$ nm was studied. Once more, there are only two propagating modes, TE$_{0}$ and TE$_{2}$, both excited with the same power. Propagation constants, $\beta_m$, can be calculated using Eq. \ref{eq:beta} and the results can be seen in Figure \ref{fig:TriWG-fields}, which presents the propagation through the MMI with total length $L_{MMI} = 200 ~ \mu m$.

\begin{figure}[ht]
\centering
\fbox{\includegraphics[width=4in]{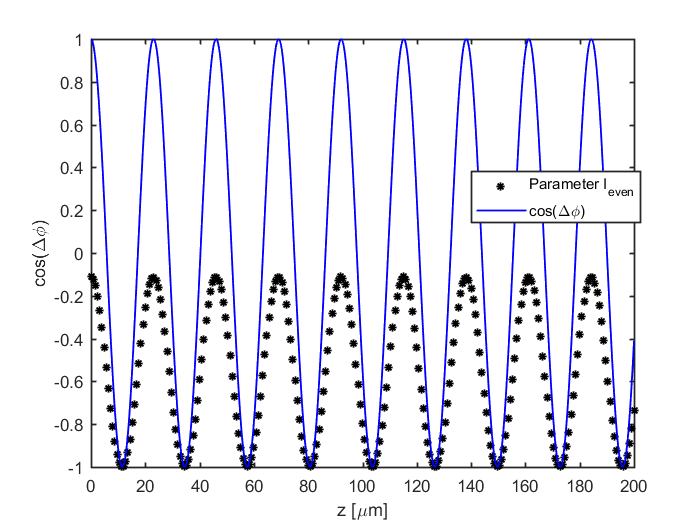}}
\caption{Comparison of the cosine of the actual phase difference ($\Delta\phi$) between the modes in the TriMW and the calculated parameter $I_{even}$.}
\label{fig:TriWG-I}
\end{figure}

According to Eq. \ref{eq:5}, the TE$_2$ mode has order $n = 2$, which yields three points with maximum electric field intensity. Measuring the intensity of the total field in these positions throughout the length of the MMI, the adimensional parameter $I_{even}$ can be calculated using Eq. \ref{eq:21}. $I_{even}$ provides a quantity that is proportional to the cosine of the phase difference between the TE$_{0}$ and TE$_{2}$ modes at a given distance. Figure \ref{fig:TriWG-I} presents both the actual cosine of the phase difference between TE$_0$ and TE$_2$ modes and the calculated $I_{even}$ parameter as a function of the propagated distance $z$. It can be observed that both curves are proportional and the value of $\Delta\phi$ can be acquired from the calculated parameter. Measuring directly a phase difference between two modes may be a complex task, but the presented method allows this information to be obtained by a simple process of measuring field intensity in $n$ specific locations, where $n$ is the order of the higher order propagating mode in the MMI. 

\section{Conclusion}

In summary, in a MMI with solely two propagating modes (fundamental and an arbitrary higher-order modes), the analytical method presented allows the calculation of the phase difference between the two modes at the end of the MMW section. The ability of measuring this phase difference, combined with recent techniques for controlled excitation of modes in MMIs, opens a new range of possible applications for this type of structure, cementing it as an important building block in the field of photonics.

\section*{Disclosure statement}
The author declares no conflict of interest.

\section*{Funding}

This work was supported by brazilian agency Fundação de Amparo à Pesquisa de Minas Gerais - FAPEMIG (under process number APQ-00822-19).

\bibliographystyle{tfnlm}
\bibliography{bibliography}

\end{document}